\documentclass[a4paper,11pt]{article}

\usepackage{jheppub}

\usepackage[abs]{overpic}

\usepackage{warpcol}

\usepackage{lineno}
\usepackage{graphicx}
\usepackage{dcolumn}
\usepackage{bm}
\usepackage{rotating}
\usepackage{epstopdf}
\usepackage{color}
\usepackage{verbatim} 
\usepackage{multirow}
\usepackage[abs]{overpic}
\usepackage{amsmath}
\usepackage{mathrsfs}
\usepackage{amssymb}
\usepackage{subfigure}
\usepackage{xspace}
\usepackage{float}
\usepackage{hyperref}
\hypersetup{colorlinks=true, linkcolor=blue, anchorcolor=blue, citecolor=blue}

\newcommand{\PreserveBackslash}[1]{\let\temp=\\#1\let\\=\temp}
\newcolumntype{C}[1]{>{\PreserveBackslash\centering}p{#1}}
\newcolumntype{R}[1]{>{\PreserveBackslash\raggedleft}p{#1}}
\newcolumntype{L}[1]{>{\PreserveBackslash\raggedright}p{#1}}

\usepackage[T1]{fontenc}

\RequirePackage{lineno}
\usepackage{epstopdf}

\title{\boldmath Measurement of the phase between strong and electromagnetic amplitudes in the decay $J/\psi\to\phi\eta$}

\collaborationImg{\includegraphics[width=.12\textwidth,origin=c,angle=90]{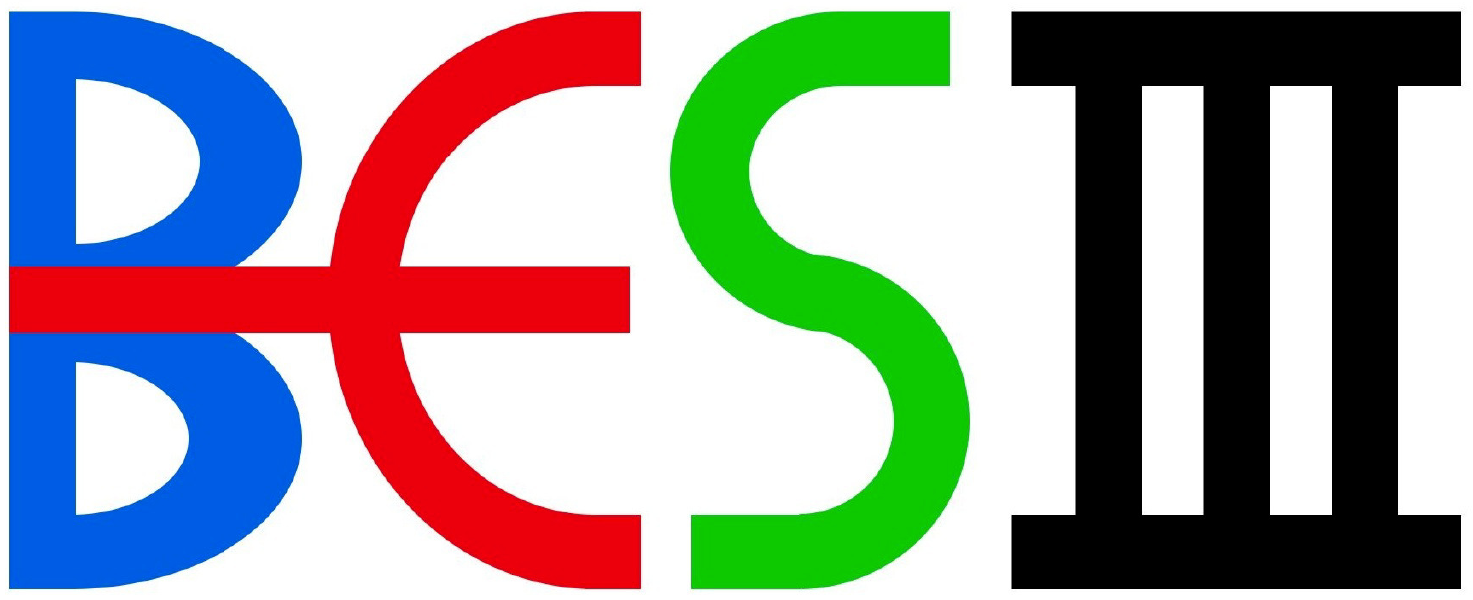}}

\collaboration{BESIII Collaboration}

\emailAdd{besiii-publications@ihep.ac.cn}

\arxivnumber{2505.05888}

\abstract{The first direct measurement of the relative phase between the strong and electromagnetic amplitudes for a $J/\psi$ decaying into a vector-pseudoscalar final state is performed using 26 energy points of $e^+e^-$ annihilation data between $3.00\ \text{GeV}$ and \mbox{3.12 GeV}.  The data sets were collected by the BESIII detector with a total integrated luminosity of 452 pb$^{-1}$. By investigating the interference pattern in the cross section lineshape of $e^+e^-\to\phi\eta$, the relative phase between the strong and electromagnetic amplitudes of $J/\psi$ decay is determined to be within $[133^\circ,228^\circ]$ at 68\% confidence level.}

\begin{document}
\maketitle
\flushbottom


\section{Introduction}

\begin{figure}[b!p]
	\centering
	\begin{minipage}[htbp]{0.25\textwidth}
		\centering
		\includegraphics[width=1\linewidth]{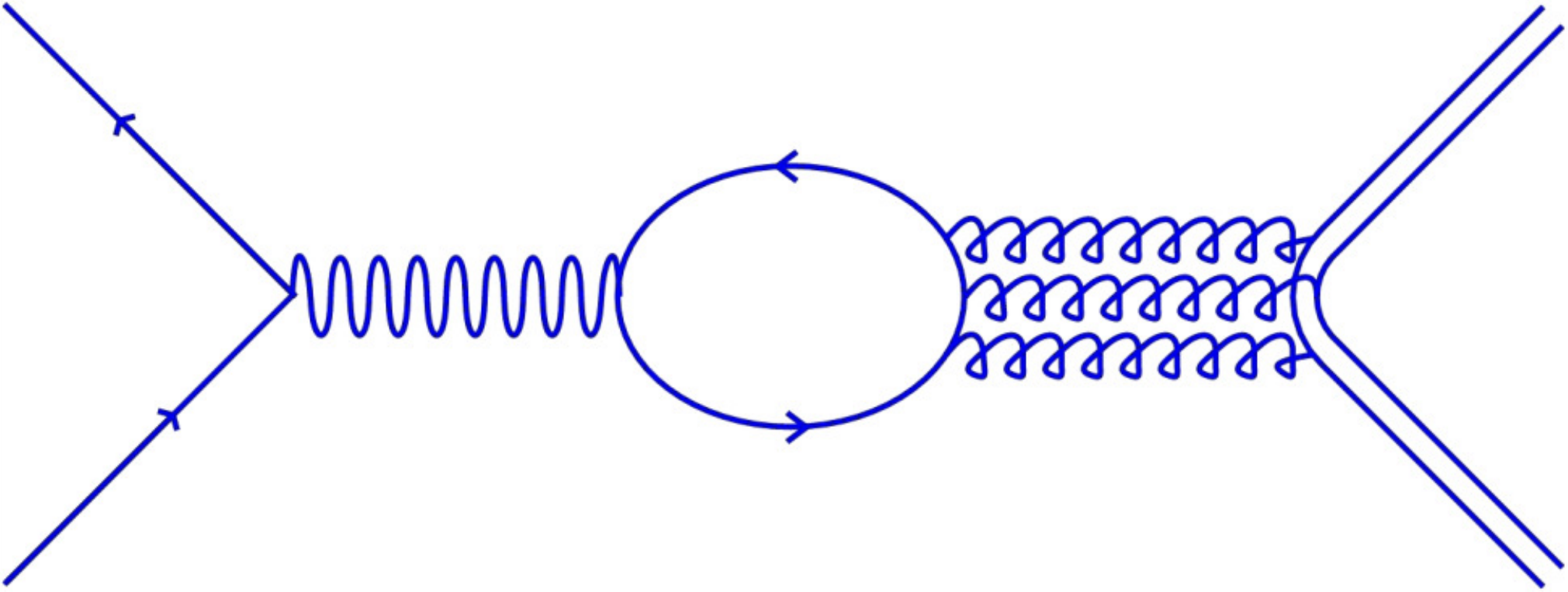}
		(a)
	\end{minipage}
    \hspace{0.05\textwidth}
	\begin{minipage}[htbp]{0.25\linewidth}
		\centering
		\includegraphics[width=1\linewidth]{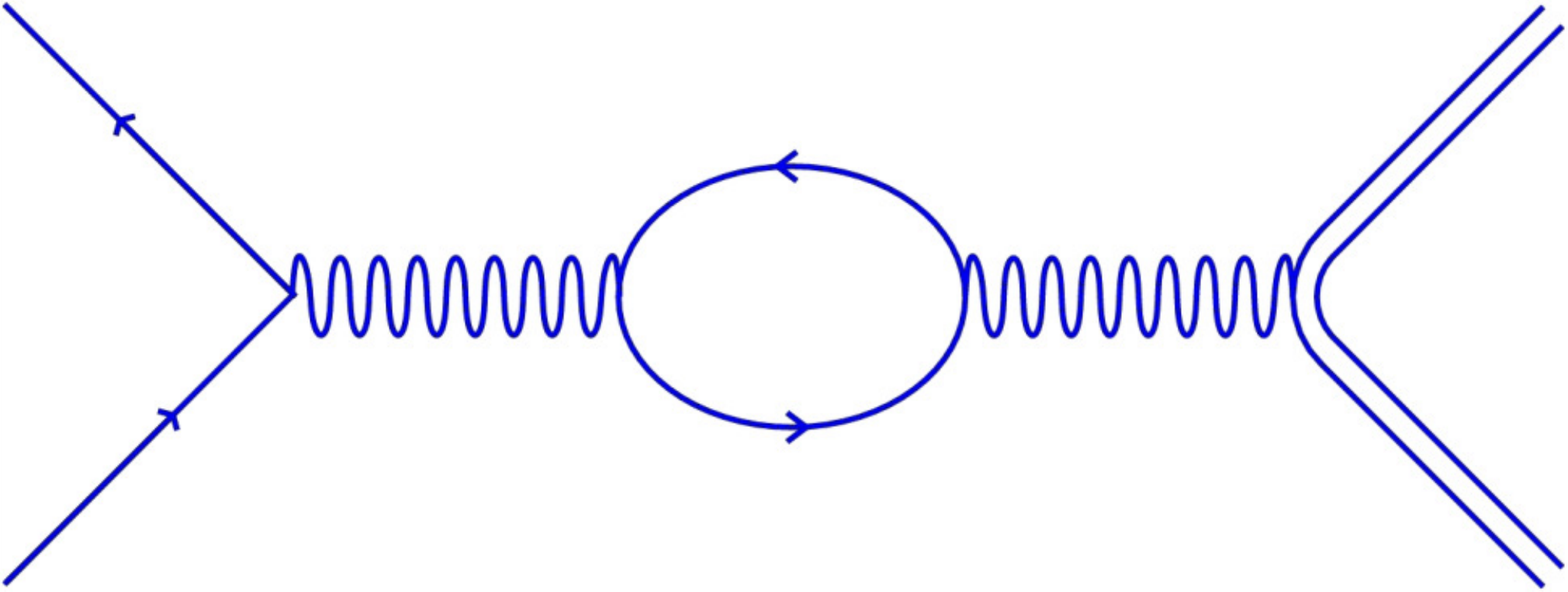}
		(b)
	\end{minipage}
    \hspace{0.05\textwidth} 
	\begin{minipage}[htbp]{0.25\linewidth}
		\centering
		\includegraphics[width=1\linewidth]{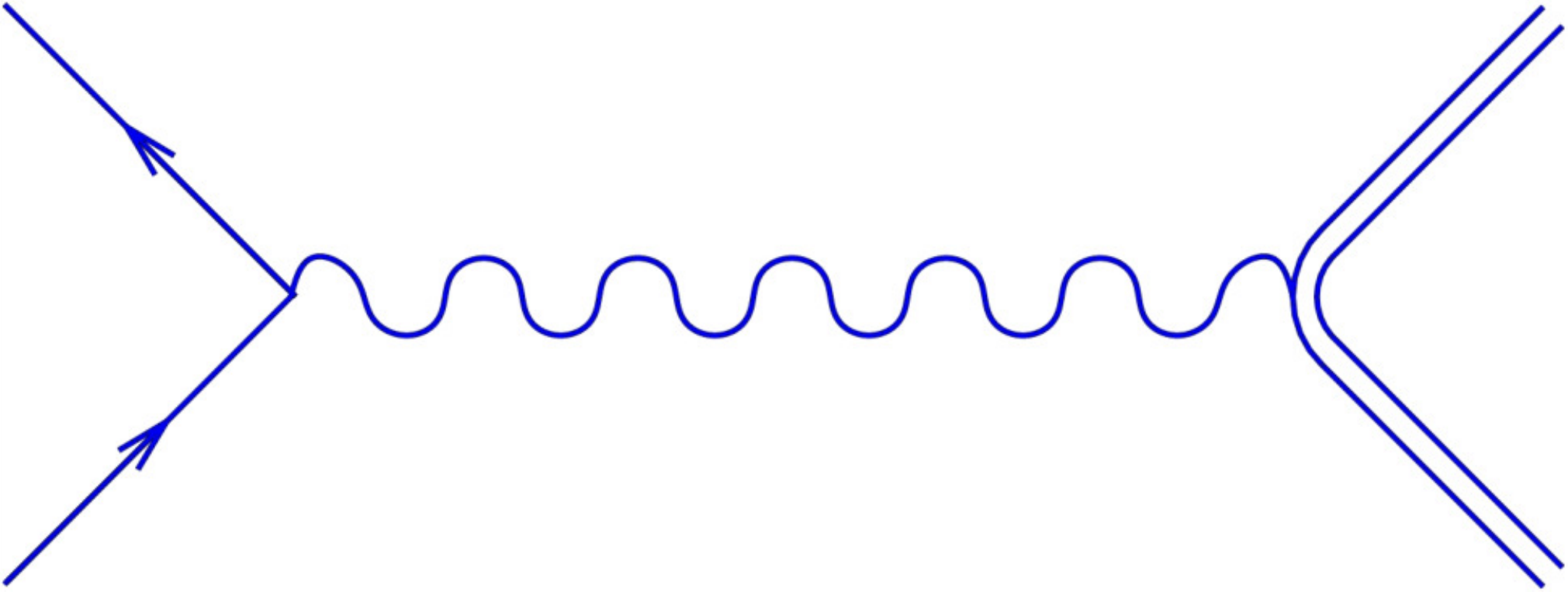}
		(c)
	\end{minipage}
	\caption{The three classes of diagrams for $e^+e^-\to\phi\eta$ in the vicinity of a charmonium \mbox{resonance \cite{BESIII:2018wid}}. The charmonium state is represented by a charm quark loop. (a) Charmonium strong decay via 3 gluons. (b) Charmonium EM decay via a virtual photon. (c) The continuum process via a virtual photon.}
	\label{Fig:sub_amplitude}
\end{figure}

At center-of-mass~(CM) energies in the vicinity of the $J/\psi$ resonance, the annihilation of $e^+e^-$ into hadronic final states can be described in terms of three amplitudes \cite{BESIII:2018wid}: $J/\psi$ production followed by the purely strong decay of the $J/\psi$ meson (mediated by 3 gluons), denoted as $\mathcal{A}_{3\mathrm{g}}$, the purely electromagnetic (EM) decay of the $J/\psi$ meson (mediated by a virtual photon), denoted as $\mathcal{A}_{\gamma}$, and the continuum Quantum Electrodynamic~(QED) process, denoted as $\mathcal{A}_{\mathrm{cont}}$, as shown in Fig. \ref{Fig:sub_amplitude}. $\mathcal{A}_{3\mathrm{g}}$ and $\mathcal{A}_{\gamma}$ proceed via $e^+e^-\to$ virtual photon $\to J/\psi\to f$, while the continuum QED process is direct $e^+e^-\to$ virtual photon $\to f$. 
The relative phase $\phi_{\gamma,\mathrm{3g}}$ between the strong and EM amplitudes for the hadronic decays of the $J/\psi$ can be directly determined by analyzing the interference pattern in the cross section lineshape of the produced particles as a function of the CM energy, $\sqrt{s}$. The total Born cross section for the $e^+e^-\to f$ process can be expressed as:
\begin{equation}
	\sigma_{f}(s) \propto |\mathcal{A}_{\mathrm{cont}}(s)+[\mathcal{A}_{\gamma}(s)+\mathcal{A}_{3\mathrm{g}}(s)\cdot e^{i\phi_{\gamma,\mathrm{3g}}}]\cdot e^{i\phi_{\mathrm{cont},\gamma}}|^2,
\end{equation}
where $\phi_{\mathrm{cont},\gamma}$ is the relative phase between EM and continuum processes, and it is determined to be $0^\circ$ by analyzing the interference patterns in the cross section lineshapes of $J/\psi\to e^+e^-$, $J/\psi\to\mu^+\mu^-$, and \mbox{$J/\psi\to\eta\pi^+\pi^-$} processes \cite{BESIII:2018wid,Boyarski:1975ci,BES:1995wyo,KEDR:2009kik}. The process of $J/\psi\to\eta\pi^+\pi^-$ violates G-parity conservation and proceeds purely through electromagnetic decay. Assuming $\phi_{\mathrm{cont},\gamma}=0^\circ$, the total cross section of $e^+e^-\to f$ in the vicinity of the $J/\psi$ resonance can be recast as:
\begin{equation}
	\sigma_{f}(s) \propto |\mathcal{A}_{\mathrm{cont}}(s)+\mathcal{A}_{\gamma}(s)+\mathcal{A}_{3\mathrm{g}}(s)\cdot e^{i\phi_{\gamma,\mathrm{3g}}}|^2.
\end{equation}

Thus far, no existing theory has provided a satisfactory explanation for the origin or implications of $\phi_{\gamma,\mathrm{3g}}$. 
Experimentally, model-dependent analyses, which rely on SU(3) flavor symmetry and symmetry breaking of light quarks, observe $\phi_{\gamma,\mathrm{3g}}$ to be around $90^\circ$ using $J/\psi$ two-body decays into meson pairs with quantum numbers ($J^P$) of $1^-0^-$ \cite{DM2:1988bfq,MARK-III:1988crp}, $0^-0^-$ \cite{Suzuki:1999nb,Kopke:1988cs}, $1^-1^-$ \cite{Kopke:1988cs}, and $1^+0^-$ \cite{Suzuki:2000yq}, and for $J/\psi$ decays into $N\bar{N}$ baryon pairs \cite{Baldini:1998en,Zhu:2015bha}. Similar model-dependent analyses suggest $\psi(2S)$ decays to pairs of mesons with $0^-0^-$ also have $\phi_{\gamma,\mathrm{3g}}$ around 90$^\circ$ \cite{CLEO:2006udi}, but $\psi(2S)$ decays to pairs of mesons with $1^-0^-$ and $1^+0^-$ are found to have a value of $\phi_{\gamma,\mathrm{3g}}$ around $0^\circ$ \cite{Suzuki:2000yq}. BESIII recently determined $\phi_{\gamma,\mathrm{3g}}$ for the $J/\psi$ in the $e^+e^-\to 5\pi$ multi-hadron process to be $(84.9\pm3.6)^\circ$ or $(-84.7\pm3.1)^\circ$, which is model independent \cite{BESIII:2018wid}. More research is needed to understand the difference between $J/\psi$ and $\psi(2S)$ decays. In addition, experimental results can be used to provide more constraints on QCD calculations.

Until now, there has been no model-independent measurement of $\phi_{\gamma,\mathrm{3g}}$ in the decay of the $J/\psi$ into vector-pseudoscalar ($VP$) mesons. The scan data collected around the $J/\psi$ resonance  by the BESIII detector provides a unique opportunity for the direct phase measurement of $J/\psi$ decays. In this analysis, we measure $\phi_{\gamma,\mathrm{3g}}$ in the process $J/\psi\to\phi\eta$ by analyzing the interference pattern in the cross section lineshape of $e^+e^-\to\phi\eta$ directly for the first time. 

\section{BESIII experiment and data sets}

The BESIII detector \cite{BESIII:2009fln} records symmetric $e^+e^-$ collisions provided by the BEPCII storage ring \cite{Yu:2016cof}, which operates with the CM energy range from \mbox{$\sqrt{s}$ = 1.85 GeV} to \mbox{4.95 GeV}, with a peak luminosity of $1.1\times10^{33}\ \mathrm{cm}^{-2}\mathrm{s}^{-1}$ achieved at $\sqrt{s}$ = 3.773 GeV. \mbox{BESIII} has collected large data samples in this energy region \cite{BESIII:2020nme,lu2020online,Zhang:2022bdc}. The cylindrical core of the \mbox{BESIII} detector covers 93\% of the full solid angle and consists of a helium-based multilayer drift chamber (MDC), a plastic scintillator time-of-flight system (TOF), and a CsI(Tl) electromagnetic calorimeter (EMC), which are all enclosed in a superconducting solenoidal magnet providing a 1.0 T magnetic field. The solenoid is supported by an octagonal flux return yoke with resistive plate counter muon-identification modules interleaved with steel. The charged-particle momentum resolution at 1 GeV/$c$ is 0.5\%, and the resolution of the rate of energy loss, d$E$/d$x$, is 6\% for electrons from Bhabha scattering. The EMC measures photon energies with a resolution of 2.5\% (5.0\%) at 1 GeV in the barrel (end-cap) region. The time resolution in the TOF barrel region is 68 ps, while that in the end-cap region is \mbox{110 ps}. The end-cap TOF system was upgraded in 2015 using multigap resistive plate chamber technology, providing a time resolution of 60 ps \cite{Li:2017jpg,Guo:2017sjt,Cao:2020ibk}.

In this analysis, the data samples collected in 2012, 2015, 2018, and 2019 at 26 different CM energies with a total integrated luminosity of about 452 pb$^{-1}$ are used. The CM energies and the integrated luminosities of each data sample are summarized in Table \ref{Tab:XSResult}. The CM energies are measured by the Beam Energy Measurement System (BEMS), in which photons from a CO$_2$ laser are Compton back scattered off the electron beam and detected by a high-purity Germanium detector \cite{Abakumova:2011rp}. The integrated luminosities are determined using $e^+e^-\to\gamma\gamma$ events \cite{BESIII:2017lkp}.

Monte Carlo (MC) simulated data samples produced with a GEANT4-based software package \cite{collaboration2003geant4}, which includes the geometric description of the BESIII detector and the detector response, are used to determine reconstruction efficiencies and to estimate backgrounds. The simulation models the beam energy spread and initial state radiation (ISR) in the $e^+e^-$ annihilation with the generator KKMC \cite{Jadach:2000ir,Jadach:1999vf}. An MC sample of $J/\psi$ inclusive decays is used to explore possible hadronic backgrounds. In this sample, the production of the $J/\psi$ resonance is simulated by the generator KKMC \cite{Jadach:2000ir,Jadach:1999vf}. The known decay modes of the $J/\psi$ are generated with \mbox{EVTGEN} \cite{Lange:2001uf,ping2008event} incorporating branching fractions from the Particle Data Group (PDG) \cite{PDG} and the remaining unknown decays are generated according to the LUNDCHARM \cite{Chen:2000tv} model. Radiation from charged final state particles is incorporated using the PHOTOS program \cite{Barberio:1993qi}. The signal MC samples for the $e^+e^-\to\phi\eta$ process at each energy point, generated using P-waves in the production process with the CONEXC generator \cite{Ping:2013jka}, which accounts for the vacuum polarization and radiative effects up to next-to-leading order, are used to estimate the reconstruction efficiency. The beam energy spread is incorporated in all MC samples. We correct the helix parameters of charged kaons to reduce the difference between simulated and data samples \cite{BESIII:2012mpj}. The $\phi$ mesons are generated with invariant masses up to 1.08 GeV/$c^2$. This range of \mbox{$0.98<M_\phi<1.08$ GeV/$c^2$} is used as a definition of the $\phi$ signal reported in this paper \cite{BESIII:2023tex}.

\section{Event selection and background analysis}

To select $e^+e^-\to\phi\eta$ events, $\phi$ and $\eta$ candidates are reconstructed through their $K^+K^-$ and $\gamma\gamma$ decay modes, respectively. Candidate events are required to have at least two candidate charged kaons with opposite charge and at least two candidate photons. 

Charged kaons detected by the MDC are required to be within the MDC acceptance of $|\cos\theta| < 0.93$, where $\theta$ is the polar angle with respect to the symmetry axis of the MDC, and their distance of closest approach to the interaction point is required to be within 10 cm along the beam direction and 1 cm in the transverse plane. For each charged track, particle identification (PID) is implemented with the specific ionization energy loss (d$E/$d$x$) measured by the MDC and the time of flight recorded by the TOF. The combined confidence levels for kaon and pion hypotheses ($\mathrm{CL}_\pi$ and $\mathrm{CL}_K$) are calculated. A kaon is identified by requiring $\mathrm{CL}_K>0.001$ and $\mathrm{CL}_K>\mathrm{CL}_\pi$.

Photon candidates are reconstructed by showers in the EMC. The photon candidates are required to be in the barrel region ($|\cos\theta|<0.80$) of the EMC with at least 25 MeV of energy deposition, and in the endcap region ($0.86<|\cos\theta|<0.92$) with at least \mbox{50 MeV} of energy deposition. To exclude showers induced by the charged tracks, the opening angle between a candidate shower and the closest charged track must be greater than $10^\circ$. To suppress electronic noise and showers unrelated to the candidate event, the difference between the EMC time and the event start time is required to be within \mbox{$[0 , 700]$ ns.}

A four-constraint (4C) kinematic fit is applied under the hypothesis $e^+e^-\to K^+K^-\gamma\gamma$, constraining the measured four-momenta of all particles to the four-momentum of the $e^+e^-$ system. For each event, the $K^+K^-\gamma\gamma$ combination with the least $\chi^2_{\mathrm{4C}}$ is retained for further study.  Events with $\chi^2_{\mathrm{4C}}>85$ are rejected.

To further suppress background, a requirement on the invariant mass of the $\gamma\gamma$ system, $M(\gamma\gamma)$, is applied, \mbox{$|M(\gamma\gamma)-M_\eta|<30\ \text{MeV}/c^2$},where $M_\eta$ is the nominal $\eta$ mass \cite{PDG} and $30$ MeV/$c^2$ corresponds to 3 times the detector resolution in the measurement of $M(\gamma\gamma)$. After applying the above selection criteria, the yield of $e^+e^-\to\phi\eta$ candidates is determined by fitting the invariant mass spectrum of the $K^+K^-$ system, $M(K^+K^-)$, in the range \mbox{$0.98<M(K^+K^-)<1.08$ GeV/$c^2$}. Potential peaking backgrounds from non-$\eta\ \phi$ processes in the $M(K^+K^-)$ spectrum are analyzed using candidate events in the sidebands of $M(\gamma\gamma)$, \mbox{$60<|M(\gamma\gamma)-M_\eta|< 90\ \mathrm{MeV}/c^2$}. An example of this background in the data sample at \mbox{$\sqrt{s} = 3096.986$ MeV} is illustrated by the blue histogram in \mbox{Fig. \ref{Fig:MkkTopo}}. It is negligible for each energy point.

\section{Observed cross section of $e^+e^-\to\phi\eta$}

The observed cross section is calculated with
\begin{equation}
    \sigma^{\mathrm{obs}}_{\phi\eta}
    =
    \frac{N_{\mathrm{sig}}}{\mathcal{L}\cdot\varepsilon\cdot\mathcal{B}}, \label{Eq:XSobsDef}
\end{equation}
where $N_{\mathrm{sig}}$ is the yield of observed signal events, $\mathcal{L}$ is the integrated luminosity, $\varepsilon$ is the detection efficiency, and $\mathcal{B}=\mathcal{B}(\phi\to K^+K^-)\cdot \mathcal{B}(\eta\to\gamma\gamma)=(19.3\pm0.2)\%$ is the product of branching fractions for the $\phi\to K^+K^-$ and $\eta\to\gamma\gamma$ decays quoted from the PDG \cite{PDG}.

\subsection{Signal yield}

\begin{figure}[htbp]
	\centering	
	
	\begin{minipage}{0.5\linewidth}
		\centering
		\includegraphics[width=1\linewidth]{"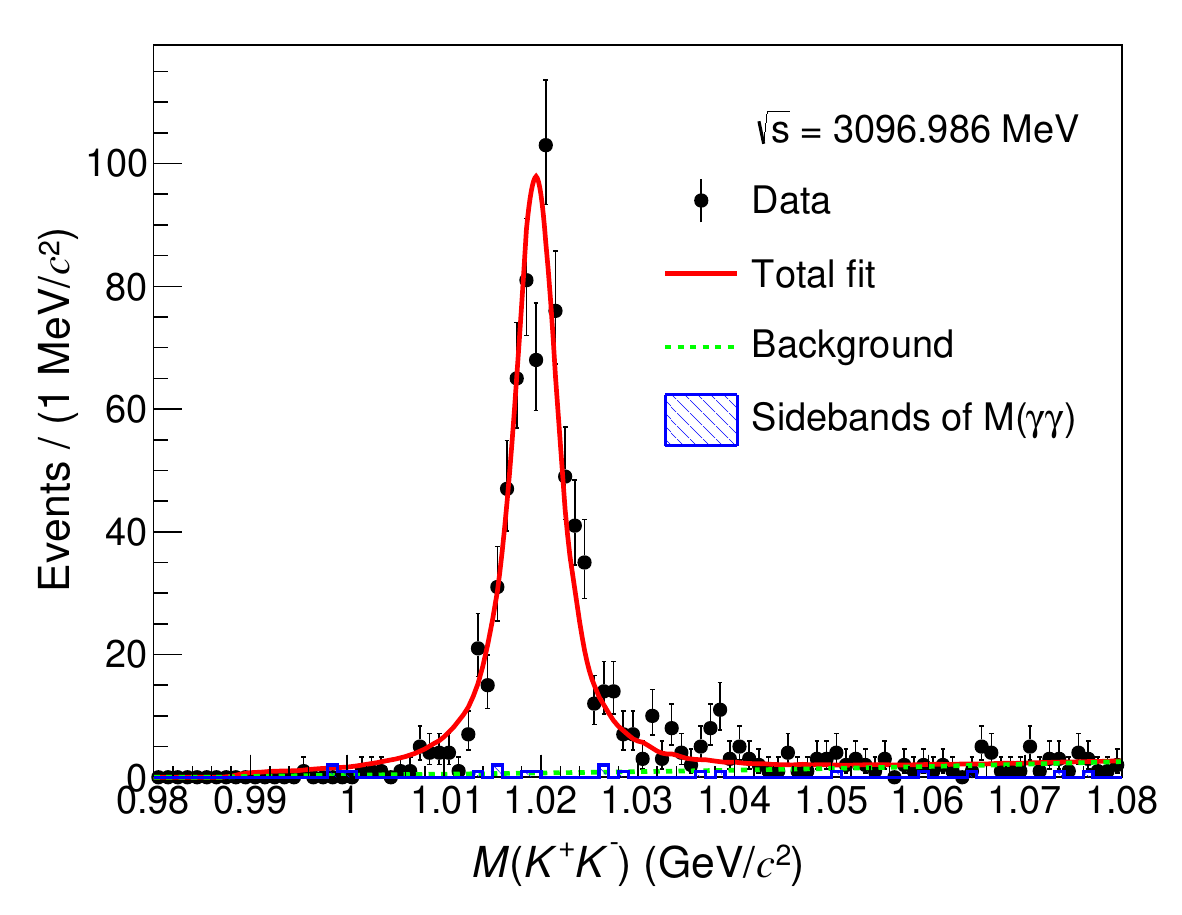"}
	\end{minipage}
	
	\caption{Fit to the $M(K^+K^-)$ distribution at $\sqrt{s}=3096.986$ MeV. The black dots with error bars are candidate events in the $\eta$ mass window of $M(\gamma\gamma)$. The red solid curve is the fit result. The green dotted line is the fitted background shape. The blue histogram is the $M(K^+K^-)$ spectrum for candidate events in the s$\eta$ sidebands of $M(\gamma\gamma)$.}
	\label{Fig:MkkTopo}
\end{figure}

The number of signal $e^+e^- \to\phi\eta$ events is determined by an unbinned maximum-likelihood fit to the $M(K^+K^-)$ spectrum. The signal is modeled by an MC-simulated shape. The background is described with an ARGUS function \cite{ARGUS:1990hfq}, with the endpoint set to the kinematic threshold of twice the $K^\pm$ mass. Due to the low statistics of the data samples, the yield follows a Poisson distribution \cite{roe2012probability} and cannot be approximated as a Gaussian distribution. Therefore, an asymmetric uncertainty is estimated. An example of the fit result for data at $\sqrt{s}=3096.986$ MeV is illustrated in Fig. \ref{Fig:MkkTopo}, and the corresponding signal yield for each energy point is presented in Table \ref{Tab:XSResult}.

\subsection{Efficiency determination} \label{sec:eff}

The reconstruction efficiency at each energy point is obtained based on the signal MC samples simulated with the CONEXC generator \cite{Ping:2013jka} and corrected by iteration. 

\begin{figure}[hthb]
	\centering
	\begin{minipage}[htbp]{0.5\linewidth}
		\centering
		\includegraphics[width=1\linewidth]{"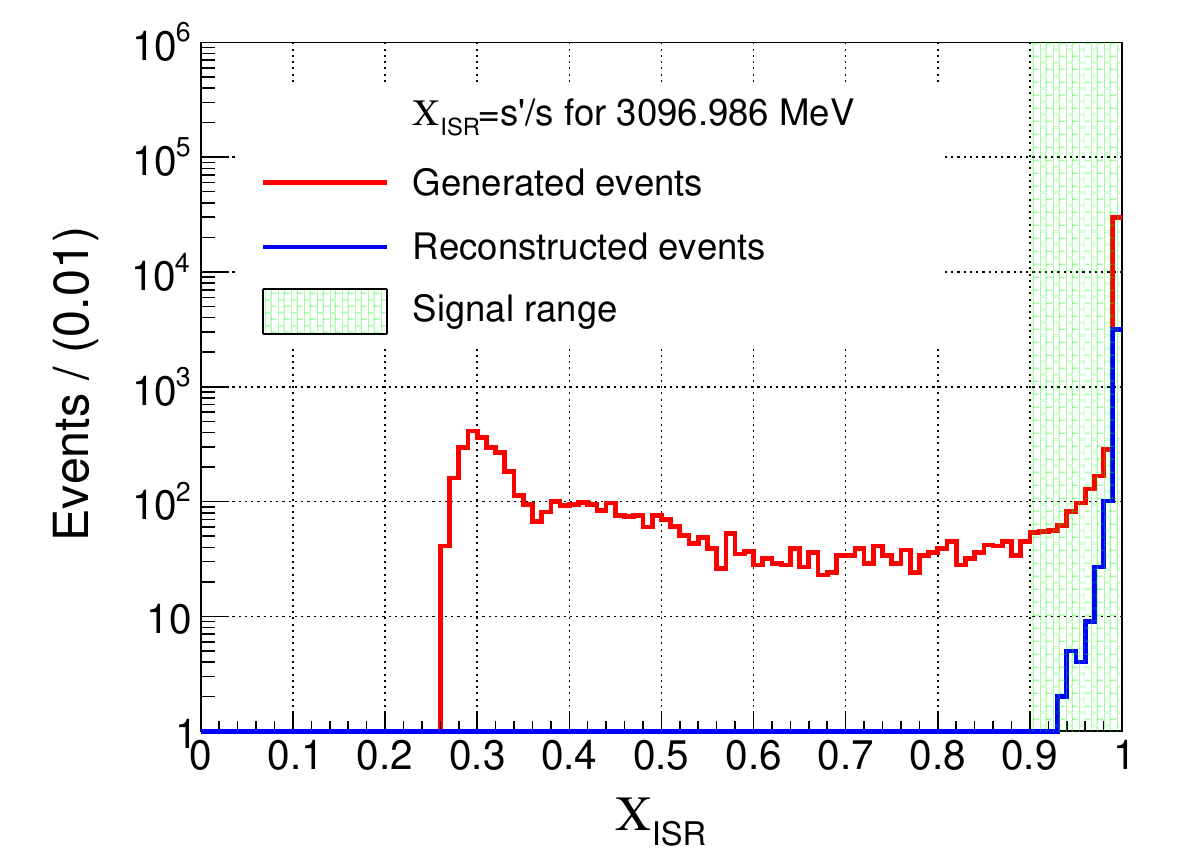"}
	\end{minipage}
	\caption{Distribution of $X_{\mathrm{ISR}}=s'/s$ in the signal MC sample at $\sqrt{s}=3096.986$~MeV. The red (blue) line shows the generated (reconstructed) events. The events in the green shaded region are used to obtain the efficiency.}
	\label{Fig:Xisr}
\end{figure}

In the CONEXC generator, the precision of simulated events with $X_{\mathrm{ISR}} = s'/s$ depends on the precision of the cross section lineshape of $e^+e^-\to\phi\eta$ used as input to the generator in the energy range below $\sqrt{s}$. Here, $s$ represents the squared energy in the CM frame of the $e^+e^-$ system before the emission of ISR photons, and $s'$ represents the squared energy after the emission. In this analysis, the cross section lineshape of $e^+e^-\to\phi\eta$ obtained by combining the measurements from BaBar \cite{BaBar:2007ceh}, BESIII \cite{BESIII:2021bjn}, and Belle \cite{Belle:2022fhh} in the range between the $\phi\eta$ mass threshold and 3.12 GeV is taken as input to the generator. Due to the deviations of the cross section lineshape among different measurements and the complexities arising from the $\phi(1680)$ and $\phi(2170)$ resonances, there is a large uncertainty on the cross section lineshape in the range between the $\phi\eta$ mass threshold and 
2.9 GeV. Figure \ref{Fig:Xisr} shows the $X_{\mathrm{ISR}}$ distribution in the MC sample simulated at \mbox{$\sqrt{s}=3096.986$ MeV}. Only events with $X_{\mathrm{ISR}}>0.9$ remain after the 4C kinematic fit. To reduce the uncertainty on the reconstruction efficiency caused by the lineshape, only the simulated events with $X_{\mathrm{ISR}}>0.9$ are used to estimate the reconstruction efficiency.

Additionally, the cross section lineshape in the range between 3.00 GeV and 3.12 GeV is not precise enough. To reasonably simulate the ISR effect, an iterative MC-generating method as described in Refs.~\cite{Ping:2013jka, Sun:2020ehv, BESIII:2021bjn, Kuraev:1985hb} is applied. The iteration procedure is repeated until the change in the cross section calculated by \mbox{Eq. \ref{Eq:XSobsDef}} is less than 0.5\%, which is the calculation uncertainty of the CONEXC generator. Finally, the reconstruction efficiency and the $\sigma^{\mathrm{obs}}_{\phi\eta}$ for each energy point are summarized in Table \ref{Tab:XSResult}.

\begin{table}[htbp]
	\centering
	 \renewcommand{\arraystretch}{1.05}
	\setlength{\belowcaptionskip}{2pt}
	\caption{Summary of the CM energy, luminosity, signal yield, efficiency, and the observed cross section of $e^+e^-\to\phi\eta$ at each energy point. Statistical uncertainties are quoted for the signal yields, efficiencies, and the observed cross sections, while both statistical and systematic uncertainties are combined in quadrature for the CM energies \cite{BESIII:2018wid} and luminosity \cite{BESIII:2018wid,BESIII:2017lkp}.}
    \label{Tab:XSResult}
    \begin{tabular}{c r@{.}l r@{.}l r@{.}l r@{.}l}
		\hline\hline
		$\sqrt{s}$ (MeV) & \multicolumn{2}{c}{$\mathcal{L}$ (pb$^{-1}$)} & \multicolumn{2}{c}{$N_{\mathrm{sig}}$} & \multicolumn{2}{c}{$\varepsilon\ (\%)$} & \multicolumn{2}{c}{$\sigma^{\mathrm{obs}}_{\phi\eta}\ (\text{pb})$} \\
		\hline
		3000.00$\pm$0.20  & 15&85$\pm$0.11  & 25&7$_{- 4.9}^{+ 5.6}$  & 36&3$\pm$0.2  &  23&1$_{-  4.4}^{+  5.0}$ \\
   		3020.00$\pm$0.20  & 17&32$\pm$0.12  & 22&0$_{- 4.3}^{+ 5.1}$  & 36&9$\pm$0.2  &  17&7$_{-  3.5}^{+  4.1}$ \\
   		3049.66$\pm$0.03  & 14&92$\pm$0.16  & 20&0$_{- 4.2}^{+ 4.8}$  & 37&3$\pm$0.2  &  18&5$_{-  3.9}^{+  4.5}$ \\
   		3058.71$\pm$0.03  & 15&06$\pm$0.16  & 28&0$_{- 5.0}^{+ 5.6}$  & 37&5$\pm$0.2  &  25&6$_{-  4.6}^{+  5.2}$ \\
    	3080.00$\pm$0.20  & 293&42$\pm$0.95 &497&3$_{-23.4}^{+24.1}$  & 37&5$\pm$0.2  &  23&3$_{-  1.1}^{+  1.2}$ \\
   		3082.51$\pm$0.04  &  4&77$\pm$0.06  &  7&1$_{- 2.4}^{+ 3.1}$  & 37&9$\pm$0.2  &  20&4$_{-  7.0}^{+  8.9}$ \\
   		3087.59$\pm$0.13  &  2&47$\pm$0.02  &  8&0$_{- 2.5}^{+ 3.2}$  & 38&1$\pm$0.2  &  44&1$_{- 13.8}^{+ 17.5}$ \\
   		3088.87$\pm$0.02  & 15&56$\pm$0.17  & 28&3$_{- 5.7}^{+ 6.4}$  & 38&0$\pm$0.2  &  24&8$_{-  5.0}^{+  5.6}$ \\
    	3091.78$\pm$0.03  & 14&91$\pm$0.16  & 35&0$_{- 5.6}^{+ 6.3}$  & 38&2$\pm$0.2  &  31&9$_{-  5.1}^{+  5.8}$ \\
    	3094.71$\pm$0.08  &  2&14$\pm$0.03  & 34&5$_{- 5.6}^{+ 6.4}$  & 39&9$\pm$0.2  & 209&8$_{- 34.1}^{+ 38.8}$ \\
    	3095.45$\pm$0.08  &  1&82$\pm$0.02  & 90&9$_{- 9.8}^{+10.2}$  & 40&1$\pm$0.2  & 645&7$_{- 70.6}^{+ 72.9}$ \\
    	3095.73$\pm$0.08  &  2&92$\pm$0.02  &291&3$_{-17.7}^{+18.4}$  & 40&4$\pm$0.2  &1277&9$_{- 79.2}^{+ 82.5}$ \\
    	3095.84$\pm$0.08  &  2&14$\pm$0.03  &328&4$_{-18.5}^{+19.1}$  & 40&2$\pm$0.2  &1979&8$_{-116.0}^{+119.8}$ \\
    	3096.20$\pm$0.07  &  4&98$\pm$0.03  &827&9$_{-31.4}^{+32.0}$  & 41&0$\pm$0.2  &2098&2$_{- 83.9}^{+ 85.4}$ \\
    	3096.99$\pm$0.08  &  3&10$\pm$0.02  &755&7$_{-29.2}^{+29.9}$  & 41&3$\pm$0.2  &3054&4$_{-124.5}^{+127.1}$ \\
    	3097.23$\pm$0.10  &  1&68$\pm$0.01  &418&1$_{-21.6}^{+22.2}$  & 41&9$\pm$0.2  &3073&3$_{-163.5}^{+168.0}$ \\
    	3097.23$\pm$0.08  &  2&07$\pm$0.03  &473&2$_{-23.3}^{+24.0}$  & 41&1$\pm$0.2  &2879&4$_{-149.9}^{+154.0}$ \\
    	3097.65$\pm$0.08  &  4&66$\pm$0.03  &860&5$_{-31.6}^{+32.2}$  & 41&8$\pm$0.2  &2286&0$_{- 88.9}^{+ 90.5}$ \\
    	3098.36$\pm$0.08  &  2&20$\pm$0.03  &229&8$_{-16.2}^{+17.1}$  & 41&1$\pm$0.2  &1316&5$_{- 95.1}^{+100.4}$ \\
    	3098.73$\pm$0.08  &  5&64$\pm$0.03  &335&1$_{-19.2}^{+20.1}$  & 41&6$\pm$0.2  & 737&2$_{- 43.1}^{+ 45.1}$ \\
    	3099.06$\pm$0.09  &  0&76$\pm$0.01  & 23&0$_{- 4.9}^{+ 5.6}$  & 41&0$\pm$0.2  & 382&4$_{- 81.7}^{+ 93.2}$ \\
    	3101.38$\pm$0.11  &  1&61$\pm$0.02  & 13&0$_{- 3.3}^{+ 4.0}$  & 41&3$\pm$0.2  & 101&0$_{- 25.5}^{+ 30.8}$ \\
    	3104.00$\pm$0.08  &  5&72$\pm$0.03  & 53&5$_{- 8.1}^{+ 8.7}$  & 41&2$\pm$0.2  & 117&6$_{- 17.9}^{+ 19.2}$ \\
    	3105.60$\pm$0.09  &  2&11$\pm$0.03  &  7&8$_{- 2.6}^{+ 3.3}$  & 40&4$\pm$0.2  &  47&4$_{- 15.9}^{+ 19.7}$ \\
    	3112.07$\pm$0.09  &  1&72$\pm$0.02  & 10&4$_{- 3.2}^{+ 3.9}$  & 39&9$\pm$0.2  &  78&3$_{- 24.1}^{+ 29.3}$ \\
    	3119.89$\pm$0.12  &  1&26$\pm$0.02  &  3&0$_{- 1.7}^{+ 2.1}$  & 38&1$\pm$0.2  &  32&2$_{- 15.2}^{+ 22.3}$ \\
		\hline\hline
	\end{tabular}
\end{table}

\subsection{Systematic uncertainty}

\begin{table}[htbp]
    \centering

    \setlength{\belowcaptionskip}{2pt}
    \caption{Relative systematic uncertainties (in percent) on the observed cross section of $e^+e^-\to\phi\eta$ at each energy point. Systematic uncertainties for each energy point arise from the efficiency, luminosity and branching fractions. The entry labeled luminosity is the luminosity systematic uncertainty, the two entries labeled $\mathcal{B}(X\to Y)$ are the branching fraction systematic uncertainties, and everything else is from the efficiency. The sources with star markers are the common and correlated systematic uncertainties. }
    \label{Tab:XsSysErr}
    \begin{tabular}{cc}
        \hline\hline
    		Source & Uncertainty (\%) \\ \hline
    		Luminosity$^*$                       & 1.0  \\
    		$\mathcal{B}(\phi\to K^+K^-)^*$      & 1.0  \\
    		$\mathcal{B}(\eta\to\gamma\gamma)^*$ & 0.5  \\
    		Tracking$^*$                         & 2.0  \\
    		PID$^*$                              & 2.0  \\
    		Photons reconstruction$^*$           & 2.0  \\
    		Kinematic fit$^*$                    & 0.3  \\
        	Iteration procedure$^*$              & 0.2  \\
    		Input lineshape$^*$                  & 1.8  \\
            Mass window of $M(\gamma\gamma)^*$   & 0.0  \\
    		$\phi$ fit range                     & 0.4  \\
    		Signal shape                         & 0.5  \\
            \hline
    		Correlated                           & 4.1  \\
            Total                                & 4.2  \\ 
        \hline\hline
    \end{tabular}
\end{table}

Several sources of systematic uncertainties are considered on the observed cross section measurement. These include differences between data and MC simulation for the tracking efficiencies, PID efficiencies, photon reconstruction efficiencies, kinematic fit, mass window selection of $M(\gamma\gamma)$, iteration procedure, input lineshape, and integrated luminosity measurement. The uncertainties from the fit procedure, and the branching fractions of the intermediate state decays are also considered.

\begin{itemize}
    \item \textit{Luminosity.} The integrated luminosity is determined using  $e^+e^-\to\gamma\gamma$ events with an uncertainty of 1.0\% \cite{BESIII:2017lkp}.

    \item \textit{Branching fractions.} The branching fractions are quoted from the \mbox{PDG \cite{PDG}}: \mbox{$\mathcal{B}(\phi\to K^+K^-)$} is $(49.1\pm0.5)$\%, with an uncertainty of 1.0\%; and $\mathcal{B}(\eta\to\gamma\gamma)$ is \mbox{($39.36\pm0.18$)\%}, with an uncertainty of 0.5\%.

    \item \textit{Tracking and PID efficiencies.} The systematic uncertainties of the tracking and PID efficiencies are both assigned as 1.0\% per track, determined using a control sample of  $e^+e^-\to K^+K^-\pi^+\pi^-$ events \cite{BESIII:2021bjn}.

    \item \textit{Photon reconstruction.} The systematic uncertainty due to the photon reconstruction is assigned to be 1.0\% per photon using a control sample of \mbox{$J/\psi\to \pi^+\pi^-\pi^0$} \mbox{events \cite{BESIII:2021bjn}.}

    \item \textit{Kinematic fit.} A helix correction is performed on the kaon tracks \cite{BESIII:2021bjn,BESIII:2012mpj} to reduce the difference between data and MC samples caused by the kinematic fit for each energy point. The difference between the reconstruction efficiency obtained from the signal MC samples with and without the helix correction is studied for each energy point. The largest deviation, 0.3\%, is taken as the systematic uncertainty for all energy points.

    \item \textit{Iteration procedure.} The systematic uncertainty associated with the iterative procedure is estimated by comparing the difference in the reconstruction efficiency between the last two iterations for each energy point. The largest, 0.2\%, is taken as the systematic uncertainty for all energy points.

    \item \textit{Input lineshape.} During the iteration mentioned in Sec. \ref{sec:eff}, the cross section lineshape of $e^+e^-\to\phi\eta$ input to the CONEXC generator is obtained by fitting the measured cross sections. The fit parameters, along with their uncertainties, obtained from the last fit iteration are listed in Table \ref{Tab:PhaseParabr}. The uncertainties on these parameters lead to an uncertainty of the efficiency obtained with the lineshape as input in the simulation of the signal MC samples. We sample 100 sets of lineshape parameters ($\phi_{\gamma,\mathrm{3g}},\mathcal{F},C$) using the Gaussian Copula method \cite{noh2009reliability} by considering their correlations. With these 100 lineshapes as inputs, 100 sets of the signal MC samples at each energy point are generated. The relative change of efficiency, $\frac{\varepsilon_i - \varepsilon}{\varepsilon}$, is calculated at each energy point, where $\varepsilon$ is the nominal efficiency and $\varepsilon_i$ is the efficiency obtained from the \(i\)\textsuperscript{th} ($i=1,2,...,100$) signal MC sample. The $\frac{\varepsilon_i - \varepsilon}{\varepsilon}$ distribution is fitted using a Gaussian function for each energy point. Conservatively, the maximum standard deviation of the Gaussian functions, 1.8\%, is taken as the systematic uncertainty on the reconstruction efficiency caused by the input lineshape for all energy points.

    \item \textit{Mass window of $M(\gamma\gamma)$.} The resolution of the mass of the $\eta$ peak is determined by fitting the $M(\gamma\gamma)$ spectrum. It is found to be 10 MeV/$c^2$ and is consistent between data and signal MC samples at each energy point. Therefore, the systematic uncertainty caused by the $\eta$ mass window in $M(\gamma\gamma)$ is ignored.

    \item \textit{Fit procedure.} The following three aspects are considered when evaluating the systematic uncertainty associated with the fit procedure. (1) \textit{$\phi$ fit range.}  To study the uncertainty caused by the $\phi$ fit range, an alternative fit is performed for each energy point by changing the $\phi$ fit range from (0.98,1.08) GeV/$c^2$ to (0.98,1.07) GeV/$c^2$. (2) \textit{Signal shape.} To study the uncertainty caused by the signal shape, an alternative fit using the MC-simulated shape convolved with a Gaussian function with free parameters is performed for each energy point. (3) \textit{Background shape.} To study the uncertainty caused by the background shape, an alternative fit using an ARGUS function with a floating endpoint is performed for each energy point. For each aspect, the largest change of the signal yield among all energy points is taken as the systematic uncertainty for all energy points.
    
\end{itemize}

A summary of all systematic uncertainties is presented in \mbox{Table \ref{Tab:XsSysErr}}. The sources marked with stars are common and correlated systematic uncertainties for different energy points and are from the efficiency, luminosity and branching fractions. The total systematic uncertainty on $\sigma^{\mathrm{obs}}_{\phi\eta}$, 4.2\%, is obtained by summing the individual uncertainties in quadrature.

\section{Cross section lineshape of $e^+e^-\to \phi\eta$} \label{sec:fitlinshape}

\begin{figure}[htbp]
	\centering	
	
	\begin{minipage}[htbp]{0.45\linewidth}
		\centering
		\includegraphics[width=1\linewidth]{"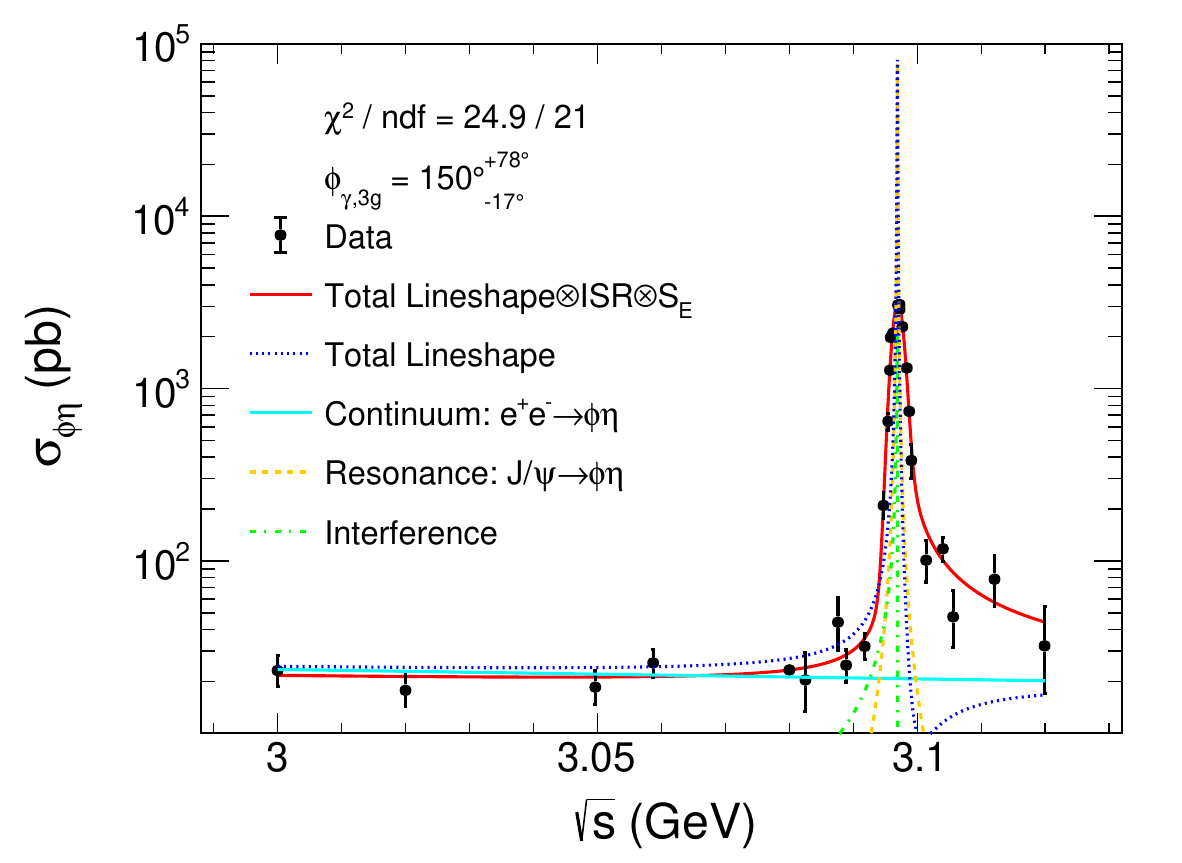"}
		(a) 
	\end{minipage}
	\begin{minipage}[htbp]{0.45\linewidth}
		\centering
		\includegraphics[width=1\linewidth]{"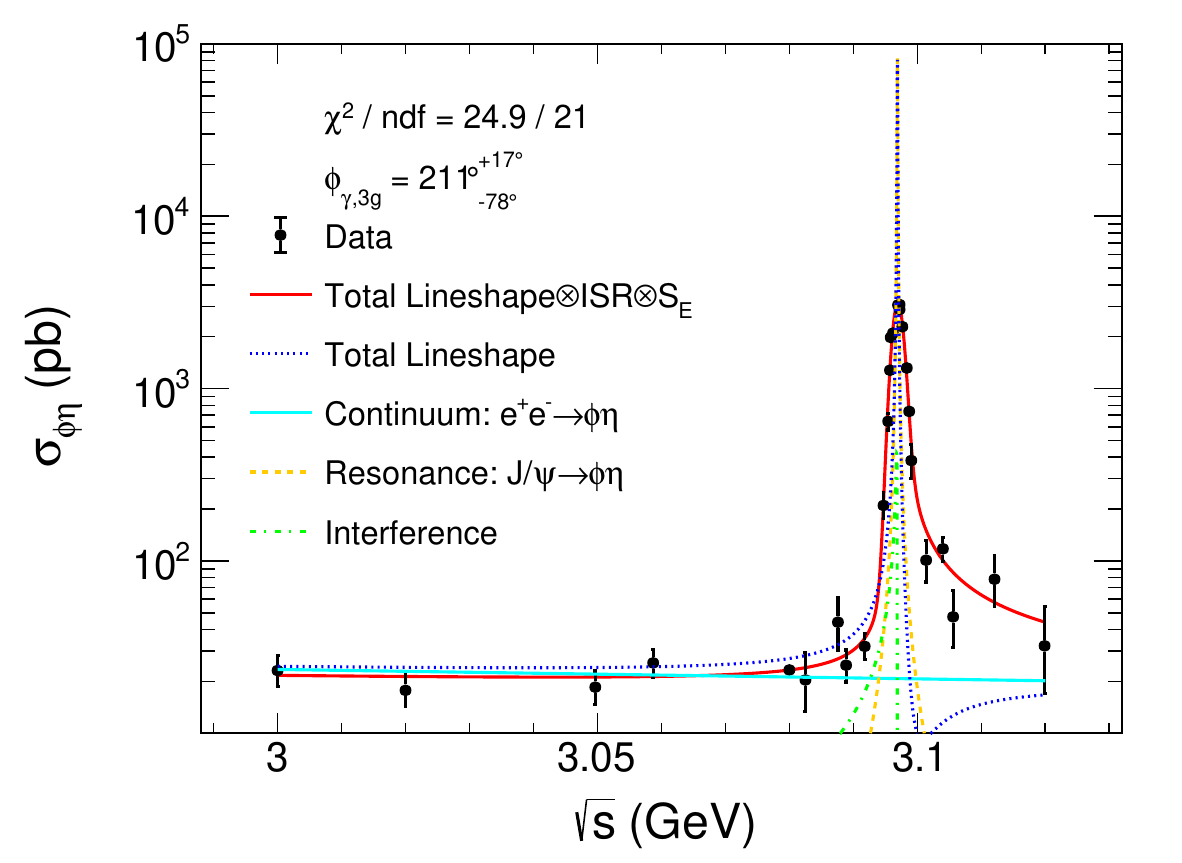"}
		(b)
	\end{minipage}

        \begin{minipage}[htbp]{0.45\linewidth}
		\centering
		\includegraphics[width=1\linewidth]{"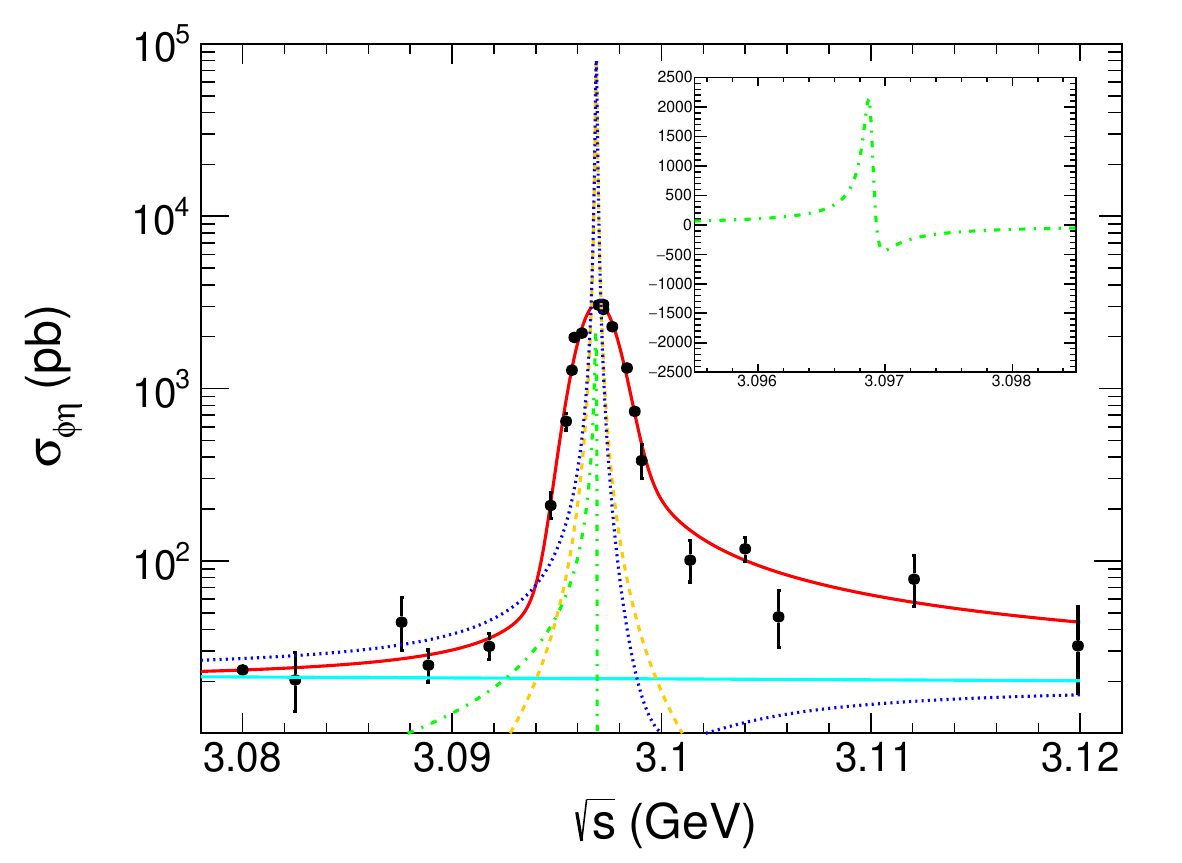"}
		(c)
	\end{minipage}
	\begin{minipage}[htbp]{0.45\linewidth}
		\centering
		\includegraphics[width=1\linewidth]{"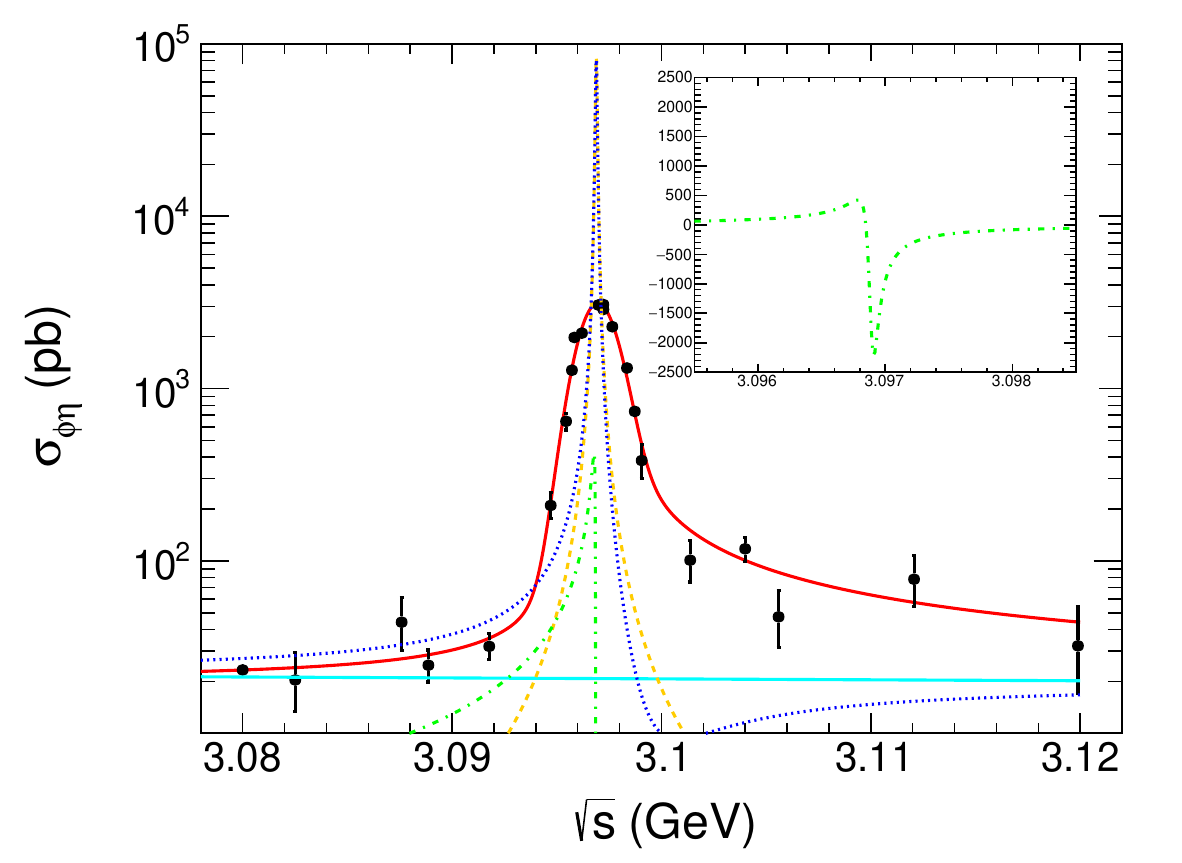"}
		(d)
	\end{minipage}
	
	\caption{Fit results of the observed cross section lineshape for $e^+e^-\to\phi\eta$. The left plot is for the positive phase of $\phi_{\gamma,\mathrm{3g}}$ and the right for the negative. The black points with error bars are the observed cross section of $e^+e^-\to\phi\eta$ at each energy point. The solid red curve denotes the overall fit result considering the effects of ISR and beam energy spread. The green solid curve is the lineshape without these two effects. The other curves show the individual contribution of each components without these two effects. The plots (c) and (d) zoom in around the $J/\psi$ resonance peak, providing a more detailed view of the interference patterns in the cross section lineshape for the positive and negative phases, respectively.} 
	\label{Fig:lineshapefit}
\end{figure}

The Born cross section of $e^+e^-\to\phi\eta$ in the vicinity of the $J/\psi$ resonance, consisting of the continuum and $J/\psi$ resonance contributions, is expressed as \cite{BESIII:2018wid,BaBar:2007ceh}: 
\begin{equation}
        \sigma(s)
         = \mathcal{P}_{\phi\eta}(s)\cdot\left(\frac{\mathcal{F}}{s^{a_0}}\right)^2\cdot\frac{4\pi\alpha^2}{3s} 
         \cdot \left|1+\frac{3}{\alpha}\frac{s}{M}\frac{\Gamma_{ee}\cdot\left(1+C\cdot e^{i\phi_{\gamma,\mathrm{3g}}}\right)}{(s-M^2)+iM\Gamma}\right|^2. \label{Eq:BornXs}
\end{equation}
Here $\mathcal{P}_{\phi\eta}(s)$ is the phase space of the $\phi\eta$ final state expressed as:
\begin{equation}
    \mathcal{P}_{\phi\eta}(s)=\left[\frac{(s-M_\phi^2-M_\eta^2)^2-4M_\phi^2M_\eta^2}{s}\right]^{3/2} \label{Eq:PHSPPhiEta}
\end{equation}
and $\mathcal{F}/s^{a_0}$ is the form factor. We set $a_0=1.5$ based on a pQCD theoretical prediction \cite{Lu:2018obb}; $\alpha$ is the fine structure constant; $M$ and $\Gamma$ are the mass and width of the $J/\psi$ meson; $\Gamma_{ee}$ is the partial width of $J/\psi\to e^+e^-$; 
and $C$ is the ratio between $|\mathcal{A}_{3\mathrm{g}}|$ and $|\mathcal{A}_{\gamma}|$.

Due to the effects of ISR and the beam energy spread, the observed cross section cannot be directly compared with the Born cross section. To take into account these effects, a two-fold numerical integration is performed to describe the expected cross section for $e^+e^-\to\phi\eta$:
\begin{equation}
        \sigma^{\mathrm{exp}}(s)
      = \int_{\sqrt{s}-5S_E}^{\sqrt{s}+5S_E}\mathrm{d}\sqrt{s'} \mathrm{G}(\sqrt{s'},S_E)\int_0^{1-X_{\mathrm{ISR}}}\mathrm{d}xF(\sqrt{s'},x)\cdot\sigma(s'\cdot(1-x)), \label{Eq:FitObsXS}
\end{equation}
where $\sqrt{s'}$ is an integration parameter with the dimension of energy; and $\mathrm{G}(\sqrt{s'},S_E)$ is a Gaussian function to describe the beam energy spread effect with a width of $S_E$. As explained in Sec. \ref{sec:eff}, $X_{\mathrm{ISR}}$ is set as $0.9$; and $x=2E_\gamma / \sqrt{s}$, where $E_{\gamma}$ is the energy of radiation photon.  The ISR function $F(W,x)$ describes the probability of ISR photon emission. From Kuraev and Fadin, it is expressed as \cite{Kuraev:1985hb}:
\begin{equation}
    F(W,x)=\beta(1+\delta)x^{\beta-1}-\beta(1-\frac{x}{2})+\frac{\beta^2}{8}\left[4(2-x)\ln\frac{1}{x}-\frac{1+3(1-x)^2}{x}\ln(1-x)-6+x\right], \label{Eq:ISRFun}
\end{equation}
with $\delta=\frac{3}{4}\beta+\frac{\alpha}{\pi}(\frac{\pi^2}{3}-\frac{1}{2})+\beta^2(\frac{9}{32}-\frac{\pi^2}{12})$ and \mbox{$\beta=\frac{2\alpha}{\pi}(2\ln\frac{\sqrt{s}}{m_e}-1)$}, where $\beta$ is the effective bremsstrahlung coupling-constant, and $m_e$ is the invariant mass of electron. We use the analytical formula given in Ref. \cite{BESIII:2018wid} for the subsequent fit to improve the efficiency of the procedure.

The relative phase $\phi_{\gamma,\mathrm{3g}}$ and other parameters ($\mathcal{F},C,S_E$) are estimated with a least-$\chi^2$ fit to $\sigma^{\mathrm{obs}}_{\phi\eta}$ using the MINUIT package \cite{james1994minuit}. The $\chi^2$ is built with an effective variance-weighted least squares method, and the correlated systematic uncertainties are considered by the factored minimization method \cite{roe2012probability}. The $\chi^2$ function reads:
\begin{equation}
    \chi^2=\sum_{i=1}^{26}\frac{(\sigma^{\mathrm{obs}}_{\phi\eta}(s_i)-f\cdot\sigma^{\mathrm{exp}}(s_i))^2}{(\Delta\sigma^{\mathrm{obs}}_{\phi\eta}(s_i))^2+[\frac{1}{2}(\sigma^{\mathrm{exp}}(s_{i,+})-\sigma^{\mathrm{exp}}(s_{i,-}))]^2} + \left(\frac{1-f}{\Delta f}\right)^2 + \sum_{i=1}^3\left(\frac{P_i^{\mathrm{PDG}}-P_i^{\mathrm{fit}}}{\Delta P_i^{\mathrm{PDG}}}\right)^2. \label{Eq:ChisqFun}
\end{equation}
In the first term of Eq. \ref{Eq:ChisqFun}, $\Delta\sigma^{\mathrm{obs}}_{\phi\eta}$ is the combined statistical and uncorrelated systematic uncertainties on the $\sigma^{\mathrm{obs}}_{\phi\eta}$ measurement, and $s_{i,\pm}=(\sqrt{s}_i\pm\Delta \sqrt{s_i})^2$, where $\Delta\sqrt{s_i}$ is the uncertainty of the CM energy measured by the BEMS, and $f$ is a normalization factor introduced as a free parameter to consider the fluctuation on the $\sigma^{\mathrm{obs}}_{\phi\eta}$ measurement caused by the correlated systematic uncertainty $\Delta f$ in Table \ref{Tab:XsSysErr}. In the third term, \mbox{$P_{i}\ (i =1,2,3)$} represents the parameters for the mass, width, and partial width of the $e^+e^-$ decay mode of the $J/\psi$ meson. These parameters are constrained by considering their uncertainties $\Delta P_{i}^{\mathrm{PDG}}$ cited from the PDG \cite{PDG}.

\begin{figure}[htbp]
	\centering	
	\begin{minipage}[htbp]{0.5\linewidth}
		\centering
		\includegraphics[width=1\linewidth]{"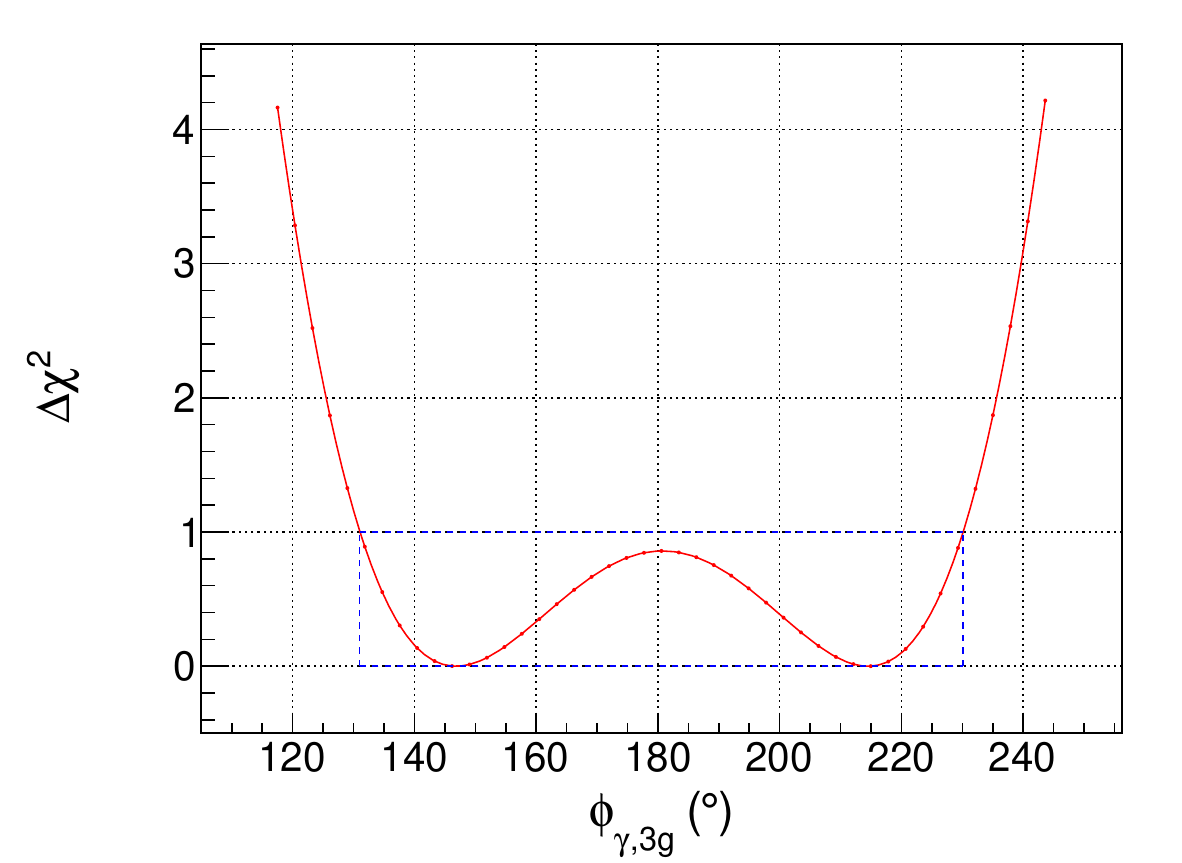"}
	\end{minipage}
	\caption{1D $\chi^2$-scan over a range of different values for $\phi_{\gamma,\mathrm{3g}}$. The dashed blue box represents the interval where $\Delta\chi^2=\chi^2-\chi^2_{\mathrm{min}}=1$, which corresponds to a $1\sigma$ confidence interval.}
	\label{Fig:1DScan}
\end{figure}

Two separate solutions with positive and negative phases $\phi_{\gamma,\mathrm{3g}}$ are found, as shown in Fig. \ref{Fig:lineshapefit}. The fitted parameters are listed in Table \ref{Tab:PhaseParabr}. The uncertainty of the fit result includes both statistical and systematic uncertainties, because all sources of systematic uncertainty have been considered in the $\chi^2$ function. The scanned $\chi^2$ curve is shown in Fig. \ref{Fig:1DScan}. As shown in Fig. \ref{Fig:1DScan}, these two solutions are indistinguishable within the $1\sigma$ confidence interval as shown in the dashed-blue box. This results from the non-linear nature of the $\chi^2$ function arising from the low statistics of the data samples. Thus, the relative phase $\phi_{\gamma,\mathrm{3g}}$ is measured to be within the range $[133^\circ,228^\circ]$ within a 1$\sigma$ confidence interval. The $S_E$ is consistent with the previous analysis reported at BESIII \cite{BESIII:2018wid}.

\begin{table}[htpb]
	\centering
	\setlength{\belowcaptionskip}{1.3pt}
	\caption{Fit results of the lineshape $e^+e^-\to\phi\eta$. The quoted uncertainties in the fit parameters include both statistical and systematic uncertainties.}
	\label{Tab:PhaseParabr}
	\begin{tabular}{lcc}
		\hline\hline
		& Positive phase & Negative phase  \\ \hline
		$\chi^2/$ndf                     & $24.9/21$     & $24.9/21$      \\
		$\phi_{\gamma,\mathrm{3g}}$ $(^\circ)$     
		& $150_{-17}^{+78}$ & $211_{-78}^{+17}$ \\
		$\mathcal{F}$ & \multicolumn{2}{c}{ 0.11$\pm$0.01} \\
		$C$                               
		& \multicolumn{2}{c}{ 3.3$\pm$0.4}   \\
		$S_E\ (\text{MeV})$                 
		& \multicolumn{2}{c}{ 0.88$\pm$0.03}  \\ 
		$f$	
		&	\multicolumn{2}{c}{ 0.99$\pm$0.04} \\
		\hline\hline
	\end{tabular}
\end{table}

\section{Summary}

For the first time, using 26 energy points of $e^+e^-$ annihilation data between $3.00\ \text{GeV}$ and 3.12 GeV, with a total integrated luminosity of \mbox{452 pb$^{-1}$}, the relative phase between strong and EM amplitudes in the decay \mbox{$J/\psi\to VP$} is measured directly through an analysis of the cross section lineshape for $e^+e^-\to\phi\eta$. The $\phi_{\gamma,\mathrm{3g}}$ for $J/\psi\to \phi \eta$ is determined to be within $[133^\circ,228^\circ]$ at the 68\% confidence level. 

\acknowledgments
The BESIII Collaboration thanks the staff of BEPCII and the IHEP computing center for their strong support. This work is supported in part by National Key R\&D Program of China under Contracts Nos. 2020YFA0406300, 2020YFA0406400, 2023YFA1606000; National Natural Science Foundation of China (NSFC) under Contracts Nos. 11635010, 11735014, 11935015, 11935016, 11935018, 12025502, 12035009, 12035013, 12061131003, 12105100, 12192260, 12192261, 12192262, 12192263, 12192264, 12192265, 12221005, 12225509, 12235017, 12361141819; the Chinese Academy of Sciences (CAS) Large-Scale Scientific Facility Program; the CAS Center for Excellence in Particle Physics (CCEPP); Joint Large-Scale Scientific Facility Funds of the NSFC and CAS under Contract No. U1832207; 100 Talents Program of CAS; The Institute of Nuclear and Particle Physics (INPAC) and Shanghai Key Laboratory for Particle Physics and Cosmology; Beijing Municipal Natural Science Foundation under Contracts No. JQ22002; German Research Foundation DFG under Contracts Nos. FOR5327, GRK 2149; Istituto Nazionale di Fisica Nucleare, Italy; Knut and Alice Wallenberg Foundation under Contracts Nos. 2021.0174, 2021.0299; Ministry of Development of Turkey under Contract No. DPT2006K-120470; National Research Foundation of Korea under Contract No. NRF-2022R1A2C1092335; National Science and Technology fund of Mongolia; National Science Research and Innovation Fund (NSRF) via the Program Management Unit for Human Resources \& Institutional Development, Research and Innovation of Thailand under Contracts Nos. B16F640076, B50G670107; Polish National Science Centre under Contract No. 2019/35/O/ST2/02907; Swedish Research Council under Contract No. 2019.04595; The Swedish Foundation for International Cooperation in Research and Higher Education under Contract No. CH2018-7756; U. S. Department of Energy under Contract No. DE-FG02-05ER41374.

\clearpage

\section*{The BESIII Collaboration}
\begin{small}
M.~Ablikim$^{1}$, M.~N.~Achasov$^{4,c}$, P.~Adlarson$^{76}$, O.~Afedulidis$^{3}$, X.~C.~Ai$^{81}$, R.~Aliberti$^{35}$, A.~Amoroso$^{75A,75C}$, Y.~Bai$^{57}$, O.~Bakina$^{36}$, I.~Balossino$^{29A}$
% ...
\\
\\{\it
$^{1}$ Institute of High Energy Physics, Beijing 100049, People's Republic of China\\
% ...
$^{3}$ Bochum  Ruhr-University, D-44780 Bochum, Germany\\
$^{4}$ Budker Institute of Nuclear Physics SB RAS (BINP), Novosibirsk 630090, Russia\\
% ...
$^{29}$ INFN Sezione di Ferrara, (A)INFN Sezione di Ferrara, I-44122, Ferrara, Italy; (B)University of Ferrara,  I-44122, Ferrara, Italy\\
% ...
$^{35}$ Johannes Gutenberg University of Mainz, Johann-Joachim-Becher-Weg 45, D-55099 Mainz, Germany\\
$^{36}$ Joint Institute for Nuclear Research, 141980 Dubna, Moscow region, Russia\\
% ...
$^{57}$ Southeast University, Nanjing 211100, People's Republic of China\\
% ...
$^{75}$ University of Turin and INFN, (A)University of Turin, I-10125, Turin, Italy; (B)University of Eastern Piedmont, I-15121, Alessandria, Italy; (C)INFN, I-10125, Turin, Italy\\
$^{76}$ Uppsala University, Box 516, SE-75120 Uppsala, Sweden\\
% ...
$^{81}$ Zhengzhou University, Zhengzhou 450001, People's Republic of China\\
\vspace{0.2cm}
% ...
$^{c}$ Also at the Novosibirsk State University, Novosibirsk, 630090, Russia\\
% ...
}


\vspace{0.4cm}
\end{small}

\end{document}